\documentclass[prd,twocolumn,preprintnumbers]{revtex4-1}

\usepackage{amsmath}
\usepackage{epsfig}
\usepackage{graphicx}
\usepackage{color}
\usepackage[normalem]{ulem}
 \usepackage{url}
\usepackage[breaklinks, plainpages=false, colorlinks=true, anchorcolor=cyan, linkcolor=red, citecolor=cyan, urlcolor=magenta, bookmarks=false]{hyperref}

\usepackage[caption=false]{subfig}

\begin{document}
\renewcommand{\thefigure}{\arabic{figure}}
\setcounter{figure}{0}

 \def\I{{\rm i}}
 \def\E{{\rm e}}
 \def\D{{\rm d}}

\bibliographystyle{apsrev}

\title{TDI on the fly}

\author{Neil J. Cornish}
\affiliation{eXtreme Gravity Institute, Department of Physics, Montana State University, Bozeman, Montana 59717, USA}

\author{Tyson B. Littenberg}
\affiliation{NASA Marshall Space Flight Center, Huntsville, AL 35812, USA}

\begin{abstract} 
Space-based gravitational wave (GW) observatories, such as the future Laser Interferometer Space Antenna (LISA), employ synthetic Time Delay Interferometry (TDI) to cancel the otherwise overwhelming laser frequency noise. The phase readouts at each spacecraft are combined with a carefully designed collection of time delays that cancel the laser frequency noise. The same collection of time delays must be applied to the GW signal models used for analysis, along with geometrical factors that encode the projection of the wave polarization tensor onto the arms of the interferometer. In principle, fully generic TDI calculations require the GW signal model to be evaluated at dozens of delay times for each data sample, a process that would require tens of millions of evaluations for a year-long signal. Here, a new method is presented that cuts the computational cost by a factor of ten thousand compared to a direct implementation at the data sample cadence, while incurring no loss in accuracy. The approach works for completely general spacecraft orbits and any flavor of TDI.
\end{abstract}

\maketitle

\section{Introduction}

In contrast to ground-based laser interferometers that have fixed arm lengths, free-flying space-based interferometers such as LISA~\cite{LISA:2024hlh} will have unequal arm lengths that vary with time. A simple Michelson-type interferometry combination formed by comparing the round-trip phase difference along two arms would be overwhelmed by laser frequency noise (LFN). An elegant solution to this problem is to synthesize an equal arm interferometer by combing phase measurements with suitable time delays~\cite{Tinto:1999yr,Tinto:2002de}. The original TDI combinations, now known as TDI-1, canceled LFN for a static unequal arm interferometer. New combinations, known as TDI-1.5, were derived to account for the overall rotation of the constellation~\cite{Shaddock:2003bc,Cornish:2003tz}, followed soon after by the TDI-2 combinations~\cite{Shaddock:2003dj} which cancel LFN for observatories with time-dependent arm lengths.

To accurately cancel LFN, the phase measurements have to be reconstructed with an accuracy of $\sim 100$ ns~\cite{Shaddock:2004ua}. The tolerances for applying TDI to the GW signal models used for analysis are considerably more relaxed, and we will show that the TDI amplitude and phase modulation for any GW signal can be perfectly reconstructed with samples taken {\em days} apart. The method can be easily applied to any existing TDI response code. The basic idea is to exploit the separation of time scales between the GW period and the orbital period of the observatory. After stripping out the carrier phase from the signal, the TDI response can be computed at a cadence set by the orbital motion (days) rather than the data (seconds) or timescales needed to construct the generic TDI response (nanoseconds).

For simplicity, we illustrate the approach using TDI-1 variables, but the method works for any TDI family. As its name suggests, TDI is most easily applied to time-domain signals, but we show how it can also be used with frequency-domain models. The approach is agnostic to the GW signal type, so long as it can be written as a sum of harmonics, and works equally well for galactic binaries, massive black hole mergers, or extreme mass ratio inspirals (EMRIs). Having a sparsely sampled TDI response is especially valuable when using accelerated likelihood evaluation techniques such as heterodyning~\cite{Cornish:2010kf,Cornish:2021lje} or wavelet domain models~\cite{Cornish:2020odn} which also sparsely sample the GW amplitude and phase. More generally, the smoothly varying sparsely sampled TDI amplitude and phase can be interpolated to give the waveform at any desired time or frequency.

\section{Time Delay Amplitude and Phase}

We will illustrate our approach using the TDI-1 interferometry variables $X(t),Y(t),Z(t)$, which correspond to synthetic Michelson-style interferometry signals extracted at spacecraft $\{1,2,3\}$. The vector connecting spacecraft $j,k$ is given by
\begin{equation}\label{armvec}
{\bf n}_i(t_k) = {\bf r}_j(t_k+L_i)-{\bf r}_k(t_k)
\end{equation}
where $\{i,j,k\}$ are even permutations of $\{1,2,3\}$ and ${\bf r}_i$ is the position vector for spacecraft $i$. The orientation and length of the vector are referenced to the time $t_k$ at spacecraft $k$ where the vector starts. The lengths of these vectors are given by the transcendental equation
\begin{equation}\label{length}
L_i(t_k) = |{\bf r}_j(t_k+L_i)-{\bf r}_k(t_k)| \, .
\end{equation}
The fact that the photon path between the two spacecraft has to point to where the receiving spacecraft will be is know as ``point ahead''. 
For simplicity, we ignore point-ahead and reference all the spacecraft positions to the solar barycenter.
The TDI-1 $X(t)$ response is then given by~\cite{Tinto:2002de}
\begin{eqnarray}\label{XTDIm}
X(t) &=&  \Psi^-_2(t - \hat{\bf k} \cdot {\bf r}_1  -2L_2)- \Psi^-_2(t - \hat{\bf k} \cdot {\bf r}_3  - L_2) \nonumber \\
&&  -\Psi^+_3(t - \hat{\bf k} \cdot {\bf r}_1  -2L_3)+ \Psi^+_3(t - \hat{\bf k} \cdot {\bf r}_2  - L_3) \nonumber \\
&& +\Psi^+_2(t - \hat{\bf k} \cdot {\bf r}_3  - L_2)- \Psi^+_2(t - \hat{\bf k} \cdot {\bf r}_1 ) \nonumber \\
&& -\Psi^-_3(t - \hat{\bf k} \cdot {\bf r}_2  -L_3)+ \Psi^-_3(t - \hat{\bf k} \cdot {\bf r}_1 )   \nonumber \\ 
&&  \Psi^+_3(t - \hat{\bf k} \cdot {\bf r}_1 - 2 L_3-2L_2)  \nonumber \\
&& - \Psi^+_3(t - \hat{\bf k} \cdot {\bf r}_2 - L_3-2L_2) \nonumber \\
&& -\Psi^-_2(t - \hat{\bf k} \cdot {\bf r}_1 - 2 L_3-2L_2) \nonumber \\
&& + \Psi^-_2(t - \hat{\bf k} \cdot {\bf r}_3 - L_2-2L_3)  \nonumber \\
&& +\Psi^-_3(t - \hat{\bf k} \cdot {\bf r}_2  - L_3-2L_2) \nonumber \\ 
&& - \Psi^-_3(t - \hat{\bf k} \cdot {\bf r}_1  - 2L_2) \nonumber \\
&& -\Psi^+_2(t - \hat{\bf k} \cdot {\bf r}_3  - L_2-2L_3)\nonumber \\ 
&& + \Psi^+_2(t - \hat{\bf k} \cdot {\bf r}_1   - 2L_3)
\end{eqnarray}
Here it is understood that the ${\bf r}_i$ are functions of time, and the individual terms in the response should be interpreted as, e.g.
\begin{equation}\label{phasesm}
\Psi_3^\pm(t - \hat{\bf k} \cdot {\bf r}_1 - L_3) = \frac{\hat{\bf n}_3(t) \cdot {\bf h}(t - \hat{\bf k} \cdot {\bf r}_1(t) - L_3(t)) \cdot \hat{\bf n}_3(t)}{2(1 \pm \hat{\bf k}\cdot \hat{\bf n}_3(t))} \, .
\end{equation}
The expressions for $Y(t)$ and $Z(t)$ follow by cyclic permutation of $\{1,2,3\}$.

Now let us write the GW signal as ${\bf h}(t) =\Re\left( \sum_n {\bf A}_n(t) e^{i \Phi_n(t)}\right)$ where ${\bf A}_n(t)$ are the real amplitude tensors for each harmonic and $\Phi_n(t)$ are the phases. The key requirement is that the amplitude and phase are smooth, slowly varying functions, which is the case for signals from binary systems. Note that while the sum of terms can always be re-written in terms of a single amplitude and phase, these combined amplitudes and phases will be rapidly varying. For example, $h = a_1 \cos(\phi_1) + a_2 \cos(\phi_2) = a \cos\phi$ with $a = \sqrt{a_1^2 + 2 a_1 a_2 \cos(\phi_1-\phi_2) +a_2^2}$ and $\phi = {\rm arctan}((a_1\sin\phi_1 + a_2 \sin\phi_2)/(a_1\cos\phi_1 + a_2 \cos\phi_2))$. 

Focusing on a single term in the sum over harmonics, and dropping the subscript $n$ for notational convenience, terms in the response such as (\ref{phasesm}) can be written as
\begin{equation}
\Psi_3^\pm(t - \Delta t) = \Re \left( A_3^\pm(t-\Delta t) e^{i \Delta \Phi(t-\Delta t)} e^{i \Phi(t)} \right) \, ,
\end{equation}
where $\Delta t =  \hat{\bf k} \cdot {\bf r}_1 + L_3$,
\begin{equation}
A_3^\pm(t - \Delta t) = \frac{\hat{\bf n}_3(t) \cdot {\bf A}(t - \Delta t) \cdot \hat{\bf n}_3(t)}{2(1 \pm \hat{\bf k}\cdot \hat{\bf n}_3(t))} 
\end{equation}
and
\begin{equation}
\Delta\Phi(t - \Delta t) = \Phi(t - \Delta t) -  \Phi(t ) \, .
\end{equation}
Schematically we then have
\begin{equation}
X(t) = \Re \left( {\cal X}(t) e^{i \Phi(t)} \right) \, ,
\end{equation}
where $ {\cal X}(t) = A_{\cal X}(t) e^{i \Phi_{\cal X}(t)}$  is a complex amplitude that can be written in terms of a smoothly varying real amplitude and phase. Note that we factor out the overall phase, $e^{i \Phi(t)}$, but not the overall amplitude $A(t)$. This is to avoid division by zero for binary merger signals, where $A(t)$ goes to zero before the time-delayed copies that appear in the TDI calculation.

Computing the TDI response comes down to extracting the TDI phases $\Phi_{\cal X}, \Phi_{\cal Y}, \Phi_{\cal Z}$ and the corresponding TDI modulated amplitudes $A_{\cal X}, A_{\cal Y}, A_{\cal Z}$. The key point is that these amplitudes and phases can be extracted on a coarse time grid. For slowly evolving systems, such as galactic binaries, the signal amplitude and phase are always slowly varying functions of time and it is sufficient to sample the complex TDI amplitude on time scales  set by the orbital motion of the detector (days). For signals with variable rates of evolution, such as massive black hole binaries, the sampling rate during the early inspiral is set by the orbital motion of the detector while near merger an adaptive sampling rate is needed to capture the fast rise in the signal amplitude. 

As is clear from the above discussion, TDI is most naturally computed in the time domain, and can be easily applied to natively time domain signal models such as the Post-Newtonian inspiral of galactic binaries, the fast EMRI waveform model~\cite{Hughes:2021exa,Katz:2021yft} and time domain inspiral-merger-ringdown models such as the IMRPhenomT~\cite{Estelles:2020osj} family of waveforms (including those with higher harmonics and spin precession such as IMRPhenomTPHM~\cite{Estelles:2021gvs}). Historically, however, most GW analyses have been performed in the frequency domain, so the majority of waveform models for massive black hole binaries are given as a function of frequency, not time. It is possible to compute the TDI response using a hybrid time-frequency approach employing the stationary phase mapping $t(f) - t_c = \Psi(f)'/(2 \pi)$, where $\Psi(f)$ is the frequency domain phase. Note that we are only using the stationary phase approximation to map between time and frequency, the amplitudes and phases remain in the frequency domain. The next step is to substitute the inverse Fourier transform
\begin{equation}
{\bf h}(t) = \int \tilde{h}(f) e^{2\pi i ft } df =  \int A(f)e^{i \Psi(f)} e^{2\pi i ft } df \, .
\end{equation}
into (\ref{XTDIm}). Terms in the response that depend on time, such as the spacecraft positions, are mapped to functions of frequency via ${\bf r}_i(t) = {\bf r}_i(t(f))$. Rather than pull out the overall phase factor $\Psi(f)$ as was done in the time domain version, here we just pull out the overall factor of $e^{2\pi i ft }$, where $t$ is the Barycenter time. The end result is an expression of the form
\begin{equation}
\tilde{X}(f) = \tilde{\cal X}(f)  \tilde{h}(f) \, ,
\end{equation}
where $\tilde{\cal X}(f)$ is a complex amplitude that includes the TDI transfer functions. Note that this method for computing the TDI response in the frequency domain has a long history. It was previously used to derive the rigid adiabatic approximation response~\cite{Rubbo:2003ap}, and was used to model the TDI response for the IMRPhenomD~\cite{Khan:2015jqa} waveform model in recent LISA Data Challenge analyses~\cite{Cornish:2020vtw,Littenberg:2023xpl}. The main difference is that in those earlier works approximations were applied that allowed the complex TDI amplitude to be extracted analytically. In contrast, here the amplitude is extracted numerically, and makes no approximations outside of using the stationary phase inspired time-frequency mapping.

\section{Implementation}

\subsection{Galactic Binaries}

The simplest application is to slowly evolving galactic binaries, where the sample cadence is entirely determined by the orbital motion of the detector. It is enough to sample the TDI amplitude and phase every few days - a saving of a factor of ten thousand compared to a direct calculation at the data sample cadence. Rather than subtract the reference phase in advance, it is simplest to compute the real and imaginary parts of $X(t)$ for the complex signal ${\bf h}(t) ={\bf A}(t) e^{i \Phi(t)}$. For a galactic binary with amplitude $A$, inclination $\iota$ and polarization angle $\psi$ the real parts of the signal are given by
\begin{eqnarray}
h^+(t) &=& A \frac{1+\cos\iota^2}{2} \cos(2\psi) \cos \Phi(t)\nonumber \\
&&  - A\cos\iota \sin(2\psi) \sin\Phi(t) \nonumber \\
h^\times(t) &=& -A \frac{1+\cos\iota^2}{2} \sin(2\psi) \cos\Phi(t) \nonumber \\
&& - A\cos\iota \cos(2\psi) \sin\Phi(t)
\end{eqnarray}
The imaginary components parts are found by replacing $\Phi(t) \rightarrow \Phi(t) + \pi/2$. The Doppler phase can be read off, modulo $2\pi$, by computing
\begin{equation}\label{Xphase}
\Phi_{\cal X} = {\rm atan2}(\Im X(t), \Re X(t)) - (\Phi(t) \,{\rm mod}\, 2\pi)\, ,
\end{equation}
while the amplitude is simply
\begin{equation}\label{amp}
A_{\cal X} = \sqrt{(\Im X(t))^2 + (\Re X(t))^2} \, .
\end{equation}
While these expressions are correct, they are not well suited to interpolation since the phase has sharp discontinuities due to the phase wrapping. In addition, the amplitude can have sharp features at zero crossings. To correct for these, we undo the $2 \pi$ wraps in the phase, and check for when the amplitude passes through zero. Since (\ref{amp}) is defined to be positive, we catch zero crossing by first locating local minima. We then evaluate the second derivative about this point using a three point stencil, and compare the standard expression to those with the sign reversed for points either side of the local minimum. If reversing the sign significantly reduces the second derivative we infer that a zero crossing occurred, and apply a sign flip to the amplitude and a $\pi$ jump to the phase. 

For galactic binaries the phase can be written in terms of a Taylor expansion $\Phi(t) = 2\pi (f_0 (t -t_0) +\frac{1}{2} \dot f_0 (t-t_0)^2)+\phi_0$. The Fourier transform of the TDI response is efficiently computed by first heterodyning the phase via
$\Phi^c_{\cal X}(t)  = \Phi_{\cal X}(t) - 2 \pi f_c t$ where $f_c =p/T_{\rm obs}$ is the carrier frequency, $T_{\rm obs}$ is the observation time, and $p =  \lfloor f_0 T_{\rm obs}\rfloor-M/4$ is the integer frequency shift that places the signal at the center of the downsampled Fourier transform. Here $M$ is the number of time samples used for the heterodyned analysis. The slowly evolving heterodyne signal $X(t) = A_{\cal X}(t) \cos\Phi^c_{\cal X}(t)$ can then be coarsely sampled in time and numerically transformed to get the Fourier domain response. The heterodyning is undone by adding $p$ to the frequency bin counter from the heterodyned FFT. This approach is similar to the original frequency-domain galactic binary response model~\cite{Cornish:2007if}, but improves upon that treatment by allowing for generic spacecraft orbits with time varying arm lengths when constructing the TDI response to the signal.

\begin{figure}[htp]
\includegraphics[width=0.45\textwidth]{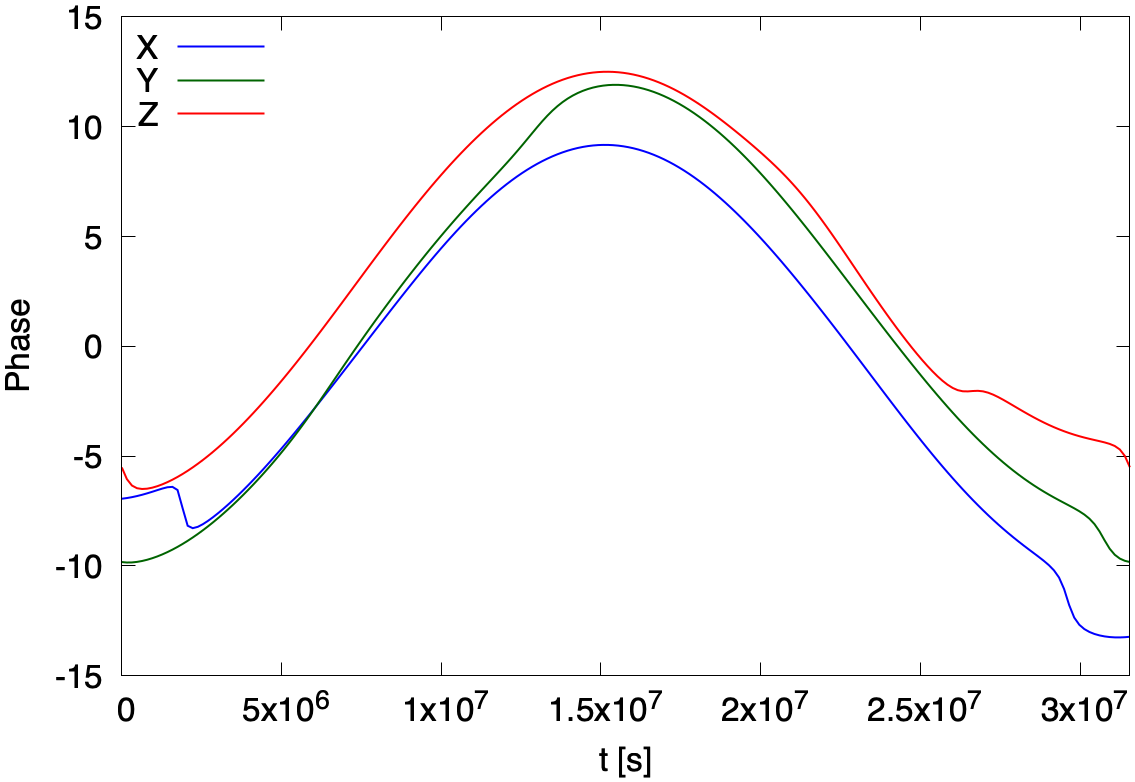} 
\includegraphics[width=0.45\textwidth]{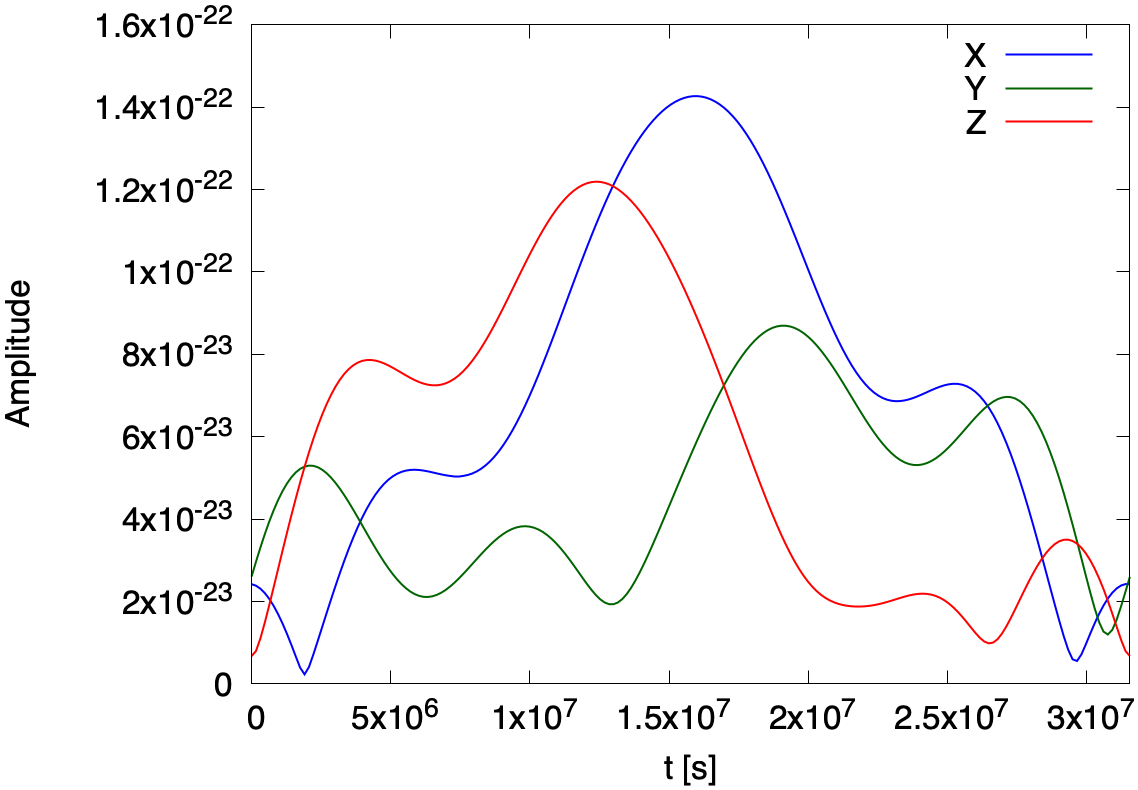} 
\caption{\label{fig:gb} The TDI phase (upper panel) and TDI amplitude (lower panel) for a galactic binary observed for one year.}
\end{figure}

Figure~\ref{fig:gb} shows the TDI amplitude and phases for a galactic binary with parameters $f_0 = 5$ mHz, $\dot f_0 = 8.15\times 10^{-16}\; {\rm Hz}^2$, $A=1.34 \times 10^{-21}$, ecliptic latitude $\alpha = -0.9$, ecliptic longitude $\beta = 3$, polarization angle $\psi = 0.8$, inclination angle $\iota = 1.5$, and initial phase $\phi_0= 1.2$. The observation time was one year. The bulk of the phase modulation is due to Doppler shifts across the orbit, while the relatively fast changes at certain times are due to the mixing of the plus and cross polarization states. The TDI calculation used $N_s=200$ time samples, and took ${\sim}0.5$ ms on a single modern CPU.

\begin{figure}[htp]
\includegraphics[width=0.45\textwidth]{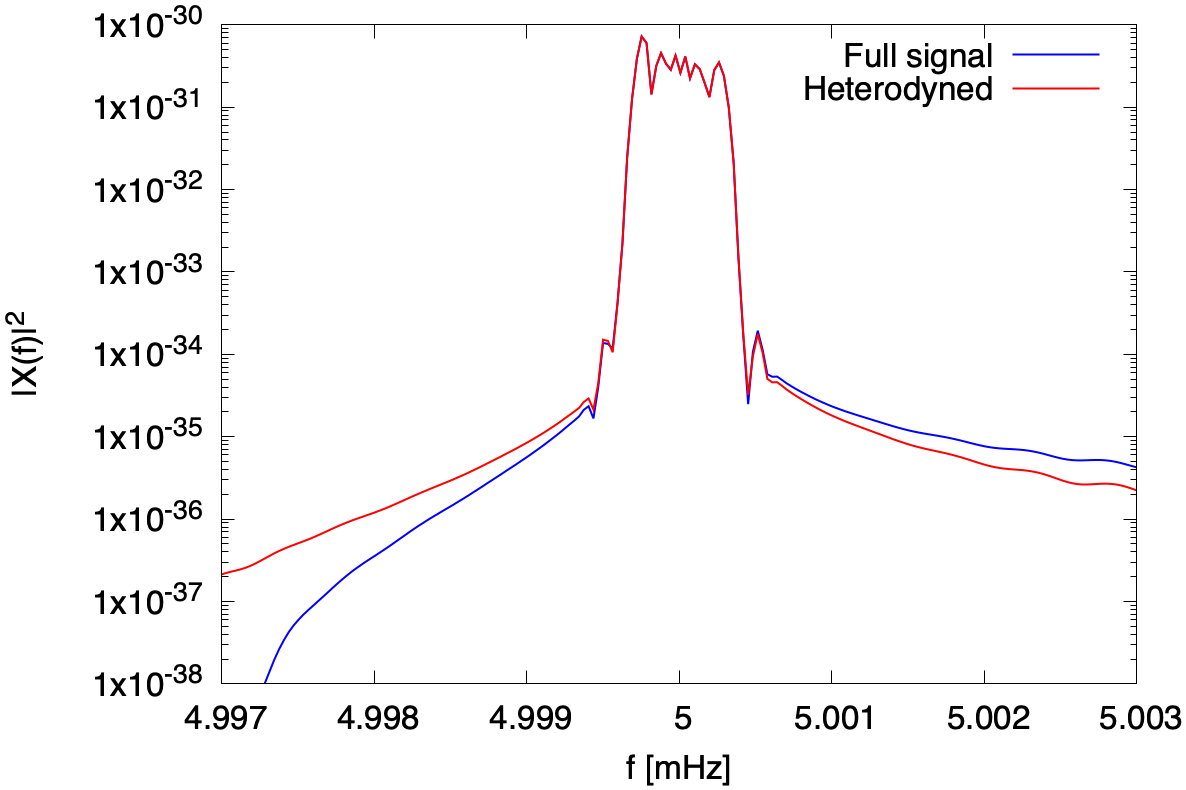} 
\caption{\label{fig:gbwave} The X-channel signal in the frequency domain computed at full cadence, and using the much faster heterodyned response.}
\end{figure}

Figure~\ref{fig:gbwave} shows the squared amplitude of the frequency domain X-channel response for the same galactic binary system computed using the full waveform and with the heterodyned waveform. The heterodyned response was computed using $M=512$ points - which is a factor of $8192$ less than for the full signal. The mismatch between the two waveforms is ${\rm MM}=1.3\times 10^{-5}$. the match can be improved by increasing $M$. 

One small technical note is that the match between the waveforms was improved by using a modified window function for the heterodyned signal. The modified window was derived by applying a forward FFT to the Tukey window function at the full sample cadence, then inverse transforming at the lower sample cadence to arrive at a slightly modified time-domain window function. Otherwise the Fourier transform of the window function, which gets convolved with the signal, differs depending on the sample cadence.

\subsection{Time domain binary black hole waveforms}

The sparse TDI response can be computed almost identically for any time-domain signal. For rapidly evolving signals, such as binary black hole mergers, the sampling rate needs to be adaptive so as to capture the rapid change in amplitude and frequency that occurs around merger, rather than at a fixed cadence set by the orbital motion of the constellation as was the case for galactic binaries. 

As an illustration of the approach, consider the IMRPhenomT model~\cite{Estelles:2020osj} , which describes the $(\ell=2,m=2)$ harmonic for spin aligned, non-pressing, quasi-circular binaries. The time spacing for the TDI computation is found by starting at merger and working forwards and backwards in time. The procedure is as follows. Starting a merger, $t=t_c$, compute the angular frequency $\omega(t)$. The time step is then $\delta t = \Delta \Phi / \omega$, where $\Delta \Phi$ is the target phase difference between $t$ and $t \pm \delta t$ (with the sign depending on whether we are working forward or backward from merger). 
We initially set $\Delta \Phi = 0.5$ radians. Once $|t-t_c| > 100 M$, where $M$ is the total mass of the system in seconds, we begin to increase the size of $\Delta \Phi$ by a factor of $\alpha$ at each time step. In the example shown below we set $\alpha = 1.1$. The time step $\delta t$ is allowed to increase until it reaches some maximum value set by the need to resolve the orbital motion of the detector, which here we set to $\delta t_{\rm max} = 2\times 10^5$ seconds, or roughly 2.3 days.

The phase and amplitude of the IMRPhenomT signal model is then evaluated on this adaptive time grid, and interpolated with a cubic spline so that they can be evaluated at the specific time samples needed in the TDI response. Because the phase evolves rapidly near merger, we compute the reference phase at one of the spacecraft rather than at the solar Barycenter, which makes unwrapping the phase difference more stable. In essence, the reference phase takes care of the large Doppler phase modulation~\cite{Cutler:1997ta}, while the remaining phase difference accounts for the TDI transfer functions and the polarization phase modulation~\cite{Cutler:1997ta}.

\begin{figure}[htp]
\includegraphics[width=0.45\textwidth]{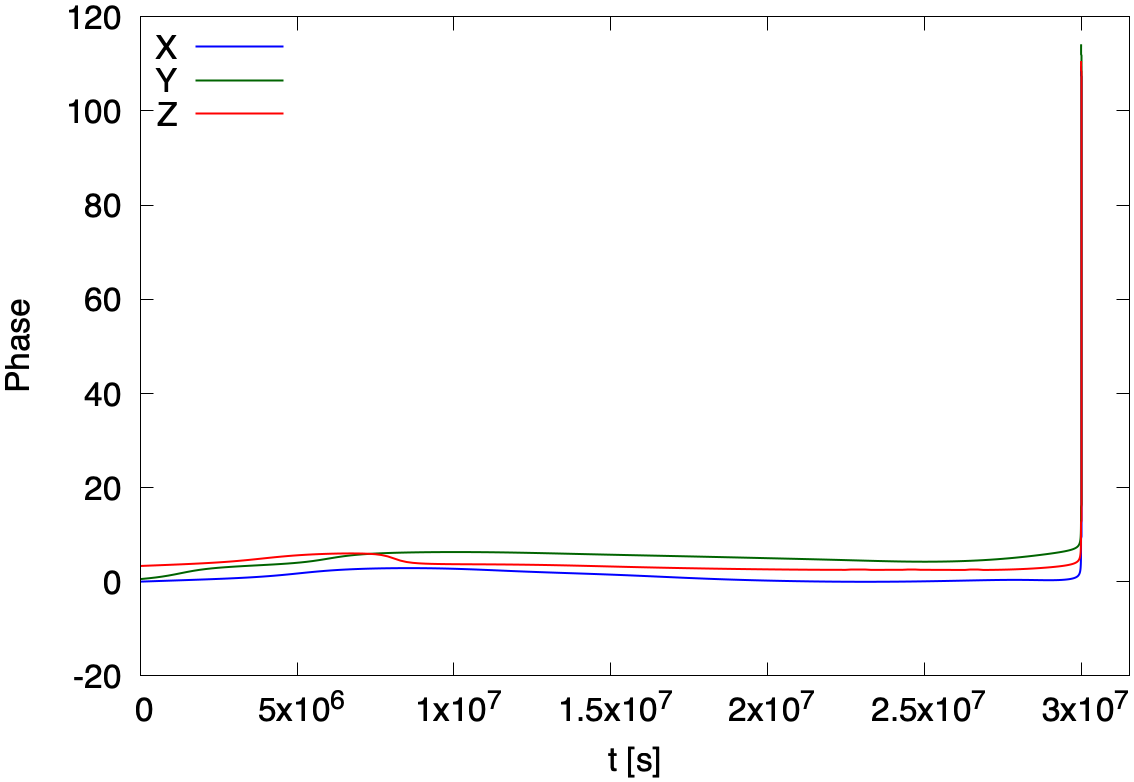} 
\includegraphics[width=0.45\textwidth]{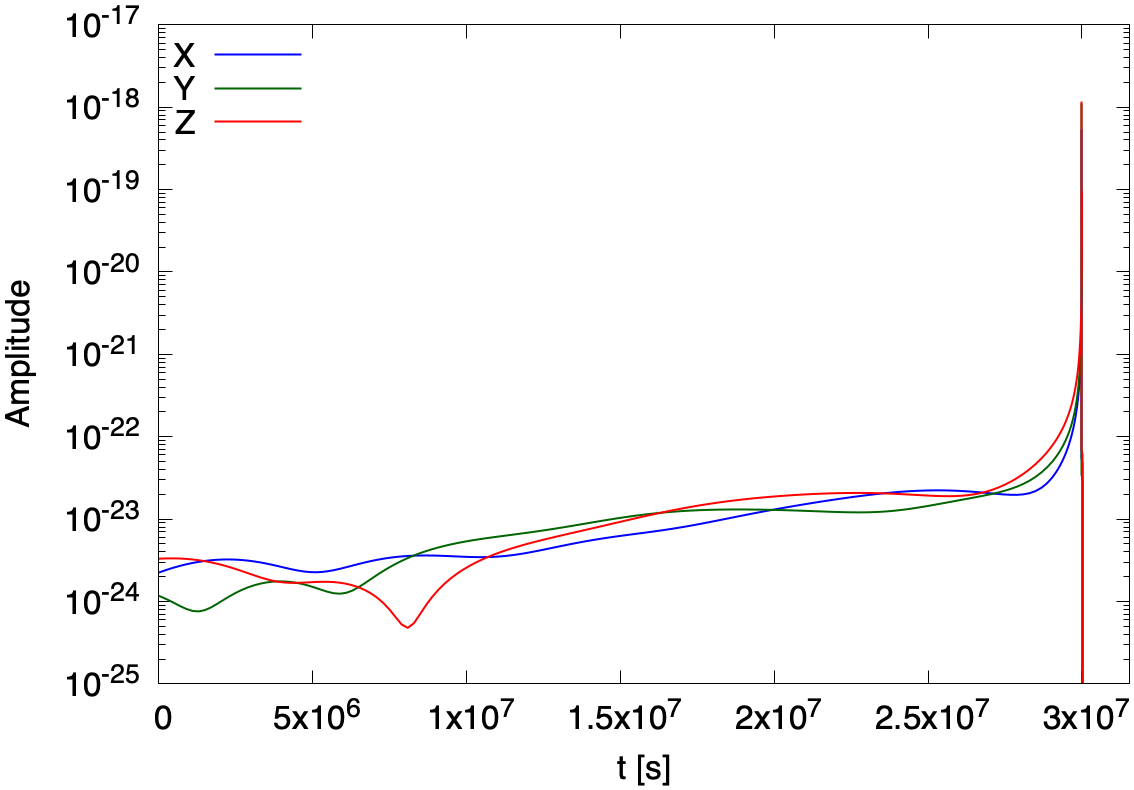} 
\caption{\label{fig:IMRPhenomT} The TDI phase (upper panel) and TDI amplitude (lower panel) for a time domain binary black hole merger}
\end{figure}

Figure~\ref{fig:IMRPhenomT} shows the TDI amplitude and phase evolution as a function of time for the IMRPhenomT waveform model. The binary parameters used in this example were as follows: detector frame masses $m_1 = 1\times 10^5 M_\odot$, $m_2 = 2\times 10^5 M_\odot$, spins $\chi_1 = 0.85$, $\chi_2 = 0.42$, merger time $3\times 10^7$ seconds, luminosity distance $D_L=1$ Gpc, merger phase $\phi_c = 0$, ecliptic co-latitude 2.31, ecliptic longitude 0.57, polarization 0.4 and cosine of inclination 0.3. The observation time was one year. The adaptive time sampling used $N_s=328$ samples, and the TDI calculation took ${\sim}1.7$ ms on a single modern CPU.

\subsection{Frequency domain binary black hole waveforms}

Frequency domain waveforms require a time-frequency mapping to be computed from the phase $\Psi(f)$. For binary black holes an adaptive frequency grid is needed to give evenly spaced time samples. A target time spacing $\Delta t$ (typically one or two days) is used to set the frequency spacing at early times, again set by the orbital motion of the spacecraft. To then determine the adaptive grid the start and end frequencies of the signal are computed. For signals that merge during the observation time the end frequency is set equal to twice the ringdown frequency, $2 f_{\rm ring}$, so as to capture the full bandwidth of the signal. Proceeding from the start frequency, the steps are given by $\delta f = \Delta t \dot f$, where $\dot f$ is computed using the leading order post-Newtonian expression. The frequency steps are constrained to be no smaller than $1/T_{\rm obs}$ (the frequency resolution of the data) and no larger than $f_{\rm ring}/\beta$. In the example below we set $\Delta t = 1\times 10^5$ seconds and $\beta = 100$.

\begin{figure}[htp]
\includegraphics[width=0.45\textwidth]{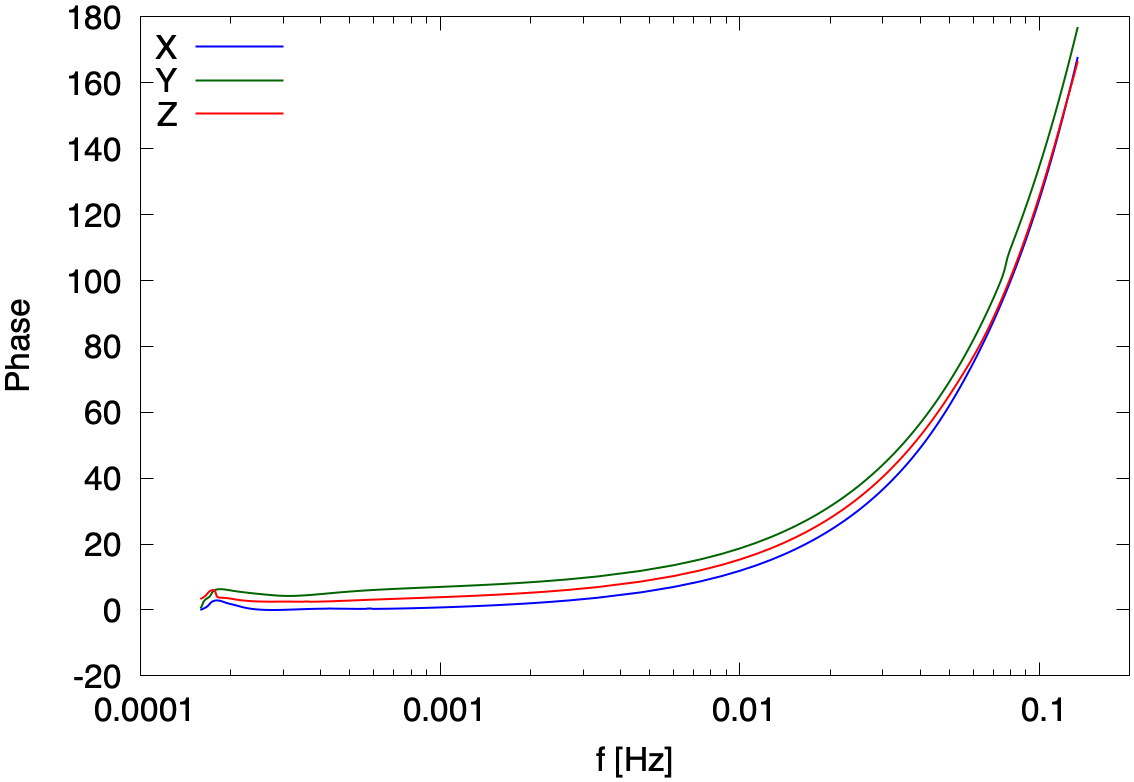} 
\includegraphics[width=0.45\textwidth]{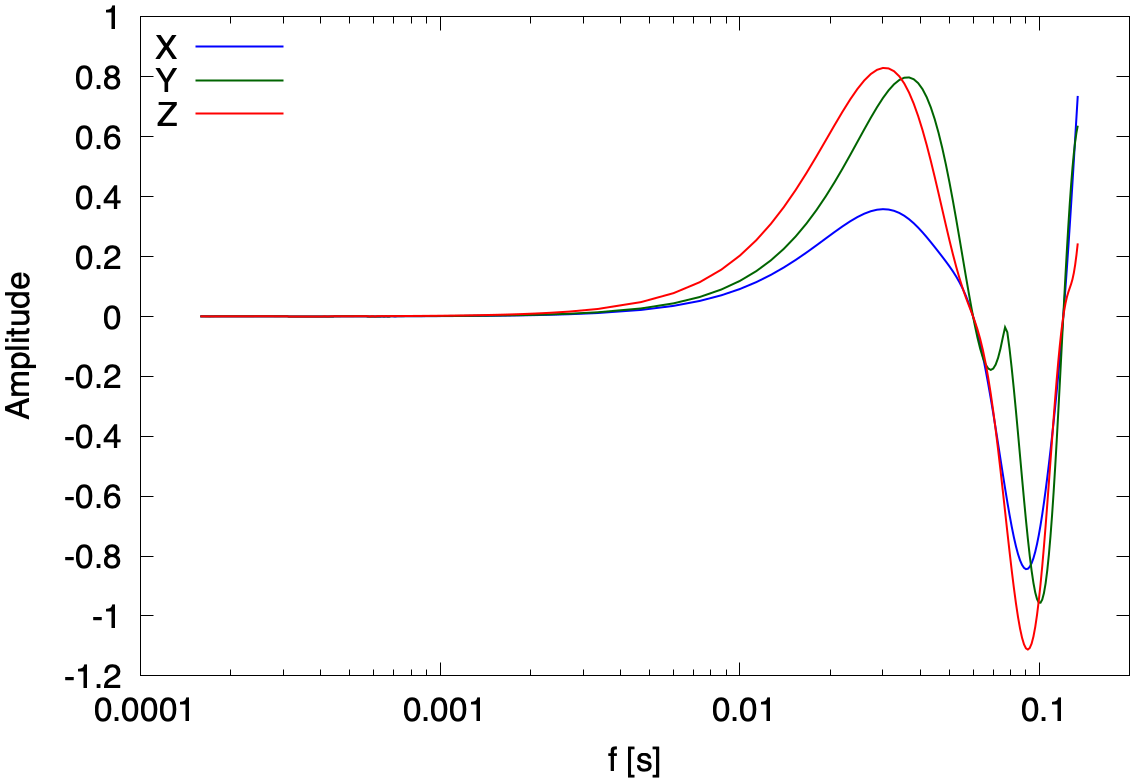} 
\caption{\label{fig:IMRPhenomD} The TDI phase (upper panel) and TDI amplitude (lower panel) for a frequency domain binary black hole merger.}
\end{figure}

Figure~\ref{fig:IMRPhenomD} shows the TDI amplitude and phase evolution as a function of time for the IMRPhenomD waveform model. The system parameters were the same as used for the IMRPhenomT time domain example. The adaptive frequency spacing required $N_s=328$ samples, and the TDI calculation took ${\sim}1.2$ ms on a single modern CPU. When using simplified constellation orbits that have equal arm lengths, the waveforms computed using this sparse TDI response exactly match those computed using the Rigid Adiabatic Approximation~\cite{Cornish:2020vtw}. The advantage of the new approach is that it can also be applied to realistic orbit models with non-constant arm lengths.

\begin{figure}[htp]
\includegraphics[width=0.45\textwidth]{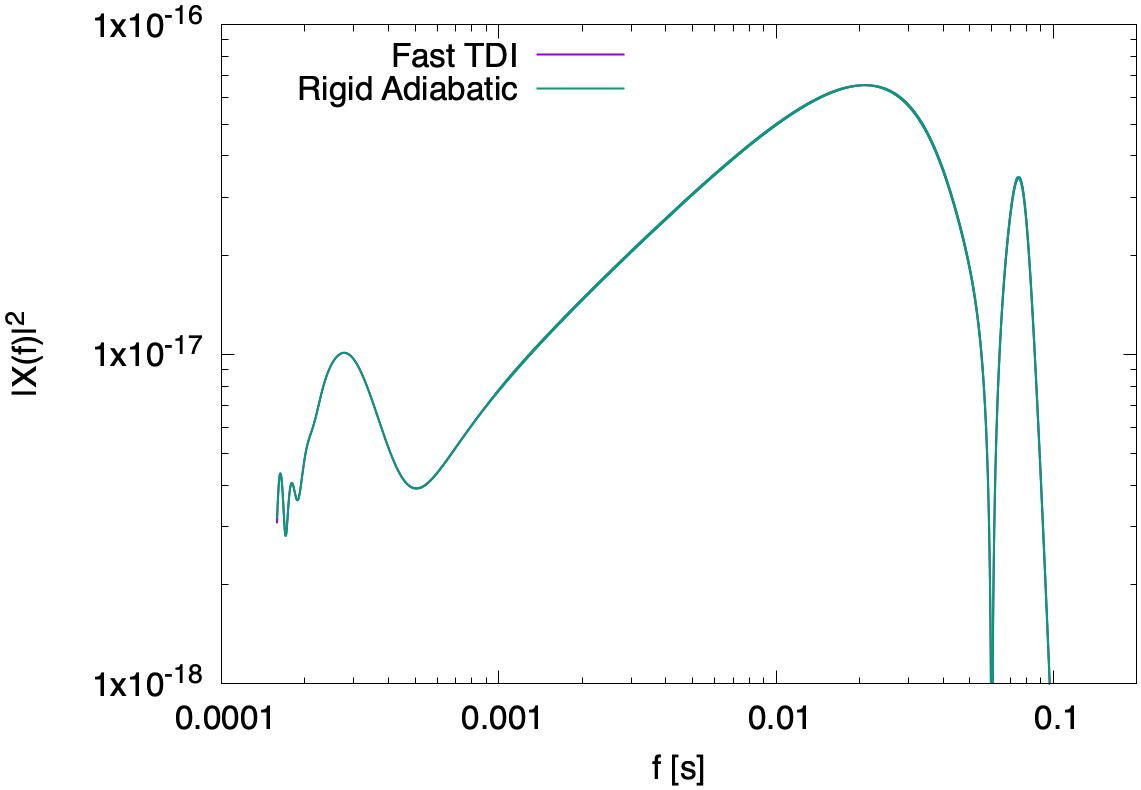} 
\caption{\label{fig:IMRPDamp} The X-channel signal computed using the new fast TDI response and the Rigid Adiabatic approximation used in past analyses.}
\end{figure}

Figure~\ref{fig:IMRPDamp} compares the X-channel Fourier domain amplitude squared computed using the new fast TDI response and the previously used Rigid Adiabatic Approximation.

\section{Discussion and Conclusions}

The sparse TDI algorithm presented here can be used with any TDI variant and with fully general spacecraft orbits. The only restriction is that the GW signals must be decomposed into a sum of harmonics. 

The combination of the sparse TDI response with a heterodyned FFT is a more accurate and general replacement to the original fast galactic binary waveform~\cite{Cornish:2007if} which has been the default model used by the LISA community for the past two decades. The adaptive TDI implementation is simpler than the recent modification of the original fast galactic binary model that accounts for time varying arm lengths~\cite{Riegger:2024qjg}.

A nice feature of the algorithm is that it can be used with any existing time-domain TDI implementation. Rather than evaluating the TDI response at every time step, the response is evaluated on a sparse grid for the sine/cosine quadratures of the signal, from which the TDI amplitude and phase are extracted.

The examples shown here cover galactic binaries and the dominant harmonic for spin aligned binary black hole models.
Work is currently in progress~\cite{harryomara} to incorporate the fast TDI response in the Fast EMRI waveform (FEW) software package~\cite{michael_l_katz_2023_8190418}.

\section*{Acknowledgments}
This work was supported by the NASA LISA Preparatory Science Grant 80NSSC24K0435 and the NASA LISA Project.

\bibliography{refs}

\end{document}